\documentclass[11pt,journal]{IEEEtran}
\usepackage{amssymb}
\usepackage{amsmath}
\usepackage[english]{babel}
\usepackage{url}
\usepackage{amsfonts}
\usepackage{graphicx}
\usepackage{epstopdf}

\graphicspath{{figures/}}

\makeatletter
\def\markboth#1#2{\def\leftmark{\@IEEEcompsoconly{\sffamily}\MakeUppercase{\protect#1}}%
\def\rightmark{\@IEEEcompsoconly{\sffamily}\MakeUppercase{\protect#2}}}
\makeatother

\begin{document}
\title{On the Performance Limits of Pilot-Based Estimation of Bandlimited Frequency-Selective Communication Channels}
\author{Francesco~Montorsi,~\IEEEmembership{Student Member,~IEEE,} and~Giorgio~Matteo~Vitetta,~\IEEEmembership{Senior~Member,~IEEE}\thanks{F.~Montorsi and G.~M.~Vitetta are with Department of Information Engineering, University of Modena e Reggio Emilia (e-mail: francesco.montorsi@unimore.it and giorgio.vitetta@unimore.it).}}
\maketitle

\begin{abstract}
In this paper the problem of assessing
bounds on the accuracy of pilot-based estimation of a bandlimited frequency
selective communication channel is tackled. \emph{Mean square error} is taken
as a figure of merit in channel estimation and a \emph{tapped-delay line
model} is adopted to represent a continuous time channel via a finite number
of unknown parameters. This allows to derive some properties of optimal
waveforms for channel sounding and closed form Cram\'{e}r-Rao bounds.
\end{abstract}

\begin{IEEEkeywords}
Estimation, Fading Channels.
\end{IEEEkeywords}

\IEEEpeerreviewmaketitle

\markboth{IEEE Transactions on Communications, accepted for publication}{F. Montorsi and G.M. Vitetta: On the Performance Limits ...}

\section{Introduction\label{sec:Introduction}}

Channel estimation plays a critical role in modern digital communication
systems, where receivers often need to acquire the channel state for each
transmitted data packet. To facilitate channel estimation, \emph{pilot}
signals, i.e. waveforms known at the receiver, are usually embedded in the
transmitted data signal \cite{tong}. In any application, it is important to
devise pilot signals in a way that, for a given figure of merit, optimality or
near optimality is ensured in a wide range of channel conditions. Important
examples of such a figure are represented by the \emph{Cram\'{e}r-Rao bound}
(CRB) and the \emph{Bayesian} CRB\ (BCRB), which limit the \emph{mean square
error} (MSE) performance achievable by any channel estimation algorithm. These
bounds have been evaluated for a pilot-aided transmission in
\emph{single-input multiple-output} (SIMO) and \emph{multiple-input
multiple-output} (MIMO) block frequency selective fading scenarios in
\cite{carvalho}, \cite{dong} under the assumptions that: a) the pilot signal
is generated by a digital modulator fed by a sequence of pilot data; b) a
symbol-spaced discrete-time model can be adopted for data transmission and, in
particular, for the representation of a multipath fading channel; c) the tap
gains of the channel model are independent and identically distributed complex
Gaussian random variables (this assumption is made in \cite{dong} only).

In this correspondence we revisit the problem of assessing performance limits
on pilot-aided channel estimation over a frequency selective channel, taking a
novel perspective. In fact, we adopt a continuous time (instead of a discrete
time\textbf{)} model for the overall description of a channel sounding system
and adopt the MSE of the estimated continuous time \emph{channel impulse
response} (CIR) as a figure of merit. Then, we show that bounds for this
figure of merit can be derived exploiting CRB's referring to the estimation of
the tap gains of a \emph{tapped delay line} (TDL) model of the communications
channel. This sheds new light on both the achieveable limits and the
properties of optimal waveforms for channel sounding; in particular, the role
played by the properties of a continuous time communication channel in
limiting the MSE performance in channel estimation is unveiled.

This Correspondence is organized as follows. In Section
\ref{sec:Signal-and-system} the model of a system for pilot-based channel
estimation is described in detail and two figures of merit for channel
estimation are defined. Two bounds on such figures are derived in Section
\ref{sec:CRB} and are evaluated in Section \ref{sec:Numerical-results} for two
different scenarios. Finally, Section \ref{sec:conclusions} offers some conclusions.

\section{Signal and System Models\label{sec:Signal-and-system}}

\begin{figure*}[tp]
\centering \includegraphics[width=6.5in]{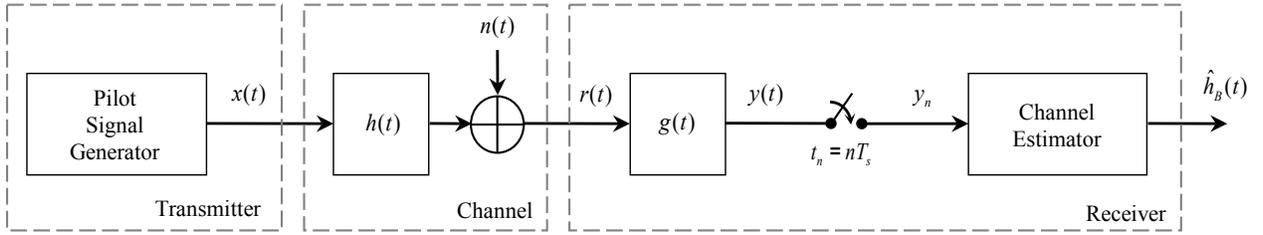} 
\caption{Channel sounding system: baseband model.}%
\label{Channel_sounder}%
\end{figure*}

In the following we consider the channel sounding system illustrated in Fig.
\ref{Channel_sounder}.
 In this system, the transmitter sends a bandlimited
real low-pass signal $x(t)$ (dubbed \emph{pilot signal} in the following),
having bandwidth $B$ and known to the receiver, over a frequency selective
communication channel characterized by its impulse response $h(t)$ (or,
equivalently, by its frequency response $H(f)$). Let $r\left(  t\right)
=x\left(  t\right)  \otimes h_{B}\left(  t\right)  +n\left(  t\right)  $
denote the noisy channel response to $x(t)$, where $\otimes$ denotes the
convolution operator, $n(t)$ is a complex circularly symmetric \emph{additive
white Gaussian noise} (AWGN) characterized by a two-sided power spectral
density $2N_{0}$ and
\begin{equation}
h_{B}\left(t\right)  \triangleq\int_{-B}^{B}H\left(  f\right)\exp\left(j2\pi ft\right)df\label{eq_4}
\end{equation}
is a bandlimited version of $h(t)$; note that $h_{B}\left(  t\right)  $ fully
describes the noiseless channel behavior in the time domain for any input
signal whose bandwidth does not exceed $B$. The noisy signal $r\left(
t\right)  $ feeds a receiver which accomplishes ideal low-pass filtering (with
bandwidth $B$), followed by sampling at a frequency $f_{s}=1/T_{s}=2B$, where
$T_{s}$ denotes the sampling period (in Fig. \ref{Channel_sounder}
$t_{n}\triangleq nT_{s}$ represents the $n$th sampling instant). We assume
that the impulse response of the low-pass filter is $g\left(  t\right)
=2B\operatorname{sinc}\left(  2Bt\right)  $, so that its frequency response
takes on a unitary value in the frequency interval $(-B,B)$; then, the filter
response $y(t)$ is given by
\begin{equation}
y(t)=x\left(  t\right)  \otimes h_{B}\left(  t\right)  +w\left(  t\right)
\text{,}\label{eq:y_def}
\end{equation}
where $w\left(  t\right)  $ is complex bandlimited Gaussian process having
zero mean and a two-sided power spectral density $S_{w}(f)=2N_{0}$ for
$\left\vert f\right\vert <B$ and zero elsewhere; note that its autocorrelation
function is $R_{w}(\tau)=4N_{0}B\operatorname{sinc}(2B\tau)$ and its average
statistical power is $\sigma_{w}^{2}=R_{w}(0)=4N_{0}B$. Sampling $y(t)$
generates the sequence $\{y_{n}\triangleq y(t_{n})\}$, which feeds a channel
estimator. This processes a finite subset of elements of $\left\{
y_{n}\right\}  $ to generate an estimate $\hat{h}_{B}\left(  t\right)  $ of
$h_{B}\left(  t\right)  $. It is important to point out that:

1. Any channel estimation algorithm assumes a specific parametric
representation of the communication channel. In the following, we adopt the
well known \emph{tapped delay line} (TDL) model for a bandlimited
communication channel \cite{bello} and assume a \emph{finite memory} (i.e., a
finite number of active taps); for this reason, $h_{B}\left(  t\right)  $ is
expressed as
\begin{equation}
h_{B}\left(  t\right)  \cong2B\sum_{l=-L_{1}}^{L_{2}}h_{B,l}%
\operatorname{sinc}\left(  2B\left(  t-\frac{l}{2B}\right)  \right)
\text{,}\label{eq_3}%
\end{equation}
where 
\begin{align*}
h_{B,l} & \triangleq \frac{1}{2B}h_{B}\left(  t_{l}\right)=\frac{1}{2B}h_{B}\left(  \frac{l}{2B}\right)\\
& =  \frac{1}{2B}\int_{-B}^{B}H\left(f\right)  \exp\left(  j2\pi l\frac{f}{2B}\right) df
\end{align*} 
for any $l$ and
$L_{1}$, ${L_{2}>0}$ (the overall number of active taps\footnote{Note that the
values of the parameters $L_{1}$\ and $L_{2}$\ (and, consequently, the value
of $L$) should be large enough to ensure a good accuracy in the representation
of the bandlimited CIR\ $h_{B}\left(  t\right)  $\ and, in particular, to
capture most of the energy of this signal. For this reason, such values mainly
depend on the power delay profile (PDP) of the considered channel and are not
necessarily equal (further details are provided in Section
\ref{sec:Numerical-results}).} is $L\triangleq L_{1}+L_{2}+1$).

2. For a given sounding waveform $x\left(  t\right)  $, a measure of the
accuracy of the channel estimate $\hat{h}_{B}\left(  t\right)  $ is provided
by the MSE, defined as
\begin{align}
\varepsilon_{B,L}&\triangleq\frac{1}{2B}\mathbb{E}_{w}\left\{  \int_{-\infty}^{+\infty}\left\vert e_{B,L}\left(  t\right)  \right\vert ^{2}dt\right\}\nonumber\\
&=\frac{1}{2B}\mathbb{E}_{w}\left\{  \int_{-B}^{B}\left\vert E_{B,L}\left(f\right)  \right\vert ^{2}df\right\}  \text{,}\label{MSE}
\end{align}
where $e_{B,L}\left(  t\right)  $ $\triangleq h_{B}\left(  t\right)  -\hat
{h}_{B}\left(  t\right)  $ and $E_{B,L}\left(  f\right)  $ $\triangleq
H_{B}\left(  f\right)  -\hat{H}_{B}\left(  f\right)  $, if the CIR
$h_{B}\left(  t\right)  $ is modelled as a deterministic unknown function, and
as
\begin{align}
\bar{\varepsilon}_{B,L}&\triangleq\frac{1}{2B}\mathbb{E}_{w,h_{B}}\left\{\int_{-\infty}^{+\infty}\left\vert e_{B,L}\left(  t\right)  \right\vert^{2}dt\right\}\nonumber\\
&=\frac{1}{2B}\mathbb{E}_{w,h_{B}}\left\{  \int_{-B}^{B}\left\vert E_{B,L}\left(  t\right)  \right\vert ^{2}df\right\}\text{,}\label{MSEbis}%
\end{align}
if $h_{B}\left(  t\right)  $ is modelled as an unknown random process. Here,
$H_{B}\left(  f\right)  $ ($\hat{H}_{B}\left(  f\right)  $)\ denotes the
Fourier continuous transform of $h_{B}\left(  t\right)  $ ($\hat{h}_{B}\left(
t\right)  $) and $\mathbb{E}_{X}\left\{  \cdot\right\}  $ denotes a
statistical average with respect to the random parameter $X$.

Substituting (\ref{eq_3}) in (\ref{eq:y_def}) yields 
\begin{equation*}
y(t)=\sum_{l=-L_{1}%
}^{L_{2}}h_{B,l}x\left(  t-\frac{l}{2B}\right)  +w\left(  t\right)\text{,}
\end{equation*}
so that the sample $y_{n}$ can be expressed as $y_{n}\triangleq y(nT_{s}%
)=y\left(  \frac{n}{2B}\right)  =\sum_{l=-L_{1}}^{L_{2}}h_{B,l}x_{n-l}%
+w_{n}$, where $x_{n}\triangleq x(t_{n})$ and $w_{n}\triangleq w(t_{n})$. In
our system model, the channel estimator processes the set of $N$ consecutive
noisy samples $\left\{  y_{n}\text{, }n=1\text{, }2\text{, }...\text{,
}N\right\}  $, i.e. the noisy vector $\mathbf{y\triangleq\lbrack}y_{1}$,
$y_{2}$, $...$, $y_{N}]^{T}$, to generate an estimate $\mathbf{\hat{h}}%
_{B}\triangleq\lbrack\hat{h}_{B,-L_{1}}$, $\hat{h}_{B,1-L_{1}}$, $...$,
$\hat{h}_{B,L_{2}}]^{T}$ of the $L$ dimensional channel parameter vector
$\mathbf{h}_{B}\triangleq\lbrack h_{B,-L_{1}}$, $h_{B,1-L_{1}}$, $...$,
$h_{B,L_{2}}]^{T}$. This results in the estimated CIR $\hat{h}_{B}\left(
t\right)  \triangleq2B\sum_{l=-L_{1}}^{L_{2}}\hat{h}_{B,l}\operatorname{sinc}%
\left(  2B\left(  t-\frac{l}{2B}\right)  \right)  $. It is easy to show that:
a) $\mathbf{y}$ can be put in matrix form as%
\begin{equation}
\mathbf{y}=\mathbf{X\,h}_{B}+\mathbf{w}\text{,}\label{eq:sig_model}%
\end{equation}
where $\mathbf{w=[}w_{1}$, $w_{2}$, $...$, $w_{N}]^{T}$ is a vector of
independent\footnote{The independence of noisy samples is due to the fact that
$\operatorname*{E}\left\{  w_{l}w_{k}^{\ast}\right\}  $ = $R_{w}(t_{l}-t_{k})$
= $4N_{0}B\operatorname{sinc}(2B(t_{l}-t_{k}))$ = $4N_{0}B\operatorname{sinc}%
(l-k)$ = $0$ if $l\neq k$. In other words, noise samples are uncorrelated and,
being jointly Gaussian random variables, are statistically independent.} and
identically distributed complex Gaussian random variables (each having zero
mean and variance $\sigma_{w}^{2}=4N_{0}B$) and $\mathbf{X}$ is a $N\times L$
matrix whose element on its $i$-th row and $j$-th column is $X_{i,j}%
=x_{i+j+L_{1}-1} $ (with $i=1$, $2$, $...$, $N$ and $j=-L_{1}$, $1-L_{1}$,
$...$, $L_{2}$); b) thanks to the property of orthogonality of the
$\operatorname{sinc}\left(  \cdot\right)  $ functions appearing in the channel
model (\ref{eq_3}), the MSE (\ref{MSE}) can be also expressed as\
\begin{equation}
\varepsilon_{B,L}=\sum_{l=-L_{1}}^{L_{2}}\mathbb{E}_{w}\left\{\left\vert h_{B,l}-\hat{h}_{B,l}\right\vert ^{2}\right\}=\sum_{l=-L_{1}}^{L_{2}}\operatorname*{MSE}\left(  \hat{h}_{B,l}\right)\text{,}\label{eq:mse}
\end{equation}
i.e. as a scaled sum of the MSE errors associated with the $L$ channel taps (a
similar expression can be developed for $\bar{\varepsilon}_{B,L}$
(\ref{MSEbis})). In the following Section the problem of deriving bounds for
the parameters $\varepsilon_{B,L}$ (\ref{MSE}) and $\bar{\varepsilon}_{B,L}$
(\ref{MSEbis}) is tackled.

\section{Evaluation of Performance Limits on Channel Estimation\label{sec:CRB}}

In estimating the vector $\mathbf{h}_{B}$ defined in previous Section, it can
be modelled as a vector of unknown \emph{deterministic} parameters or as a
vector of \emph{random} \ parameters with given statistical properties. In
this Section we take into consideration both models, deriving some new bounds
on the channel estimation accuracy.

\subsection{CRB-based performance limit}
In this Paragraph we focus on the class of \emph{unbiased} estimators of the
unknown deterministic vector $\mathbf{h}_{B}$ and derive a lower bound for the
parameter $\varepsilon_{B,L}$ (\ref{eq:mse}). To begin, we note that
$\varepsilon_{B,L}$ can be evaluated as $\varepsilon_{B,L}=\sum_{l=-L_{1}%
}^{L_{2}}\operatorname*{var}(\hat{h}_{B,l})$, since $\mathbb{E}_{w}%
\{|h_{B,l}-\hat{h}_{B,l}|^{2}\}=\operatorname*{var}(\hat{h}_{B,l})$ (with $l=$
$-L_{1}$, $1-L_{1}$, $...$, $L_{2}$), where $\operatorname*{var}(X)$ denotes
the variance of the random variable $X$. A lower bound to $\operatorname*{var}%
(\hat{h}_{B,l})$ for the above mentioned class of estimators is represented by
the CRB \cite{kay}, which, in this case, can be expressed as\footnote{Note
that, to ease the reading, the indices of the rows and of the columns of
$\mathbf{J}_{C}\left(  \mathbf{h}_{B}\right)  $ and $\mathbf{J}_{C}%
^{-1}\left(  \mathbf{h}_{B}\right)  $ range from $-L_{1}$ to $L_{2}$.}
$\operatorname*{var}(\hat{h}_{B,l})\geq\left[  \mathbf{J}_{C}^{-1}\left(
\mathbf{h}_{B}\right)  \right]  _{l,l}$ with $l=$ $-L_{1}$, $1-L_{1}$, $...$,
$L_{2}$, where

\begin{equation}
\begin{split} & \left[\mathbf{J}_{C}(\mathbf{h}_{B})\right]_{l,p}\triangleq\\
 & \quad\mathbb{E}_{\mathbf{y}}\left.\left\{ \frac{\partial\ln f_{\mathbf{y}}\left(\mathbf{y;\mathbf{\mathbf{\tilde{h}}}}_{B}\right)}{\partial\tilde{h}_{B,l}^{\ast}}\left(\frac{\partial\ln f_{\mathbf{y}}\left(\mathbf{y;\mathbf{\mathbf{\tilde{h}}}}_{B}\right)}{\partial\tilde{h}_{B,p}^{\ast}}\right)^{\ast}\right\} \right\vert _{\mathbf{\tilde{h}}_{B}=\mathbf{h}_{B}}\label{eq_8}
\end{split}
\end{equation}
with $l,\,p=$ $-L_{1}$, $1-L_{1}$, $...$, $L_{2}$, is an $L\times L$ complex matrix, known as \emph{Fisher Information Matrix}
(FIM), $f_{\mathbf{y}}\left(  \mathbf{y;\mathbf{\mathbf{h}}}_{B}\right)  $ is
the joint probability density function of $\mathbf{y}$ (\ref{eq:sig_model})
parameterized by the unknown (random) vector $\mathbf{h}_{B}$ and
$\mathbf{\mathbf{\tilde{h}}}_{B}\triangleq\lbrack\tilde{h}_{B,-L_{1}}$,
$\tilde{h}_{B,1-L_{1}}$, $...$, $\tilde{h}_{B,L_{2}}]^{T}$ is a (deterministic) trial vector\footnote{The trial vector is used to indicate that the differentiation operation in the FIM definition is against a deterministic (versus random) complex variable. In particular, if $f(\boldsymbol{\mathbf{\mu}})$ is some function of the deterministic complex vector $\boldsymbol{\mathbf{\mu}}=[\dots,\mu_i,\dots]$, then the usual definition $\frac{\partial f(\boldsymbol{\mathbf{\mu}}) }{\partial \mu_{i}^{*}}\triangleq\frac{1}{2}\left(\frac{\partial f(\boldsymbol{\mathbf{\mu}})}{\partial\mathrm{Re}\left\{ \mu_{i}\right\} }+j\frac{\partial f(\boldsymbol{\mathbf{\mu}})}{\partial\mathrm{Im}\left\{ \mu_{i}\right\} }\right)$ applies.}.
Then, the lower bound
\begin{equation}
\varepsilon_{B,L}\geq\sum_{l=-L_{1}}^{L_{2}}\left[  \mathbf{J}_{C}^{-1}\left(
\mathbf{h}_{B}\right)  \right]  _{l,l}=\operatorname{tr}(\mathbf{J}_{C}%
^{-1}\left(  \mathbf{h}_{B}\right)  )\label{eq_8bis}%
\end{equation}
can be formulated for $\varepsilon_{B,L}$, where $\operatorname{tr}%
(\mathbf{A})$ denotes the \emph{trace} of a square matrix $\mathbf{A}$. From
the model (\ref{eq:sig_model}) it can be easily inferred that, given
$\mathbf{h}_{B}=\mathbf{\mathbf{\tilde{h}}}_{B}$, $\mathbf{y}\sim
\mathcal{C}\mathcal{N}(\mathbf{\mu}$, $\mathbf{R}_{\mathbf{w}})$, where
$\mathbf{\mu\triangleq X\mathbf{\tilde{h}}}_{B}$ and $\mathbf{R}%
_{\mathbf{w}}=\sigma_{w}^{2}\mathbf{I}_{N}$ is the covariance matrix of
$\mathbf{w}$ ($\mathbf{I}_{N}$ is the $N\times N$ identity matrix), so that
the element on $l$-th row and $p $-th column of $\mathbf{J}_{C}%
(\mathbf{\mathbf{\mathbf{h}}}_{B})$\ can be expressed as (e.g. see
\cite[Paragraph 2]{delmas_abeida}, \cite[rel. B.3.25]{stoica_moses})
\begin{align}
\left[  \mathbf{J}_{C}\left(  \mathbf{h}_{B}\right)  \right]  _{l,p}=2\operatorname*{Re}\left[  \left(  \frac{\partial\mathbf{\mu}}{\partial\tilde{h}_{B,l}^{\ast}}\right)  ^{H}\mathbf{R}_{\mathbf{w}}^{-1}\frac{\partial\mathbf{\mu}}{\partial\tilde{h}_{B,p}^{\ast}}\right]_{\mathbf{\mathbf{\mathbf{\tilde{h}}}}_{B}\mathbf{\mathbf{\mathbf{=}}h}_{B}%
}\nonumber\\
+\left.\operatorname*{tr}\left(  \mathbf{R}_{\mathbf{w}}^{-1}\frac{\partial\mathbf{R}_{\mathbf{w}}}{\partial\tilde{h}_{B,l}^{\ast}}\mathbf{R}_{\mathbf{w}}^{-1}\frac{\partial\mathbf{R}_{\mathbf{w}}}{\partial\tilde{h}_{B,p}^{\ast}}\right)  \right\vert_{\mathbf{\mathbf{\mathbf{\tilde{h}}}}_{B}\mathbf{\mathbf{\mathbf{=}}h}_{B}}\label{eq_9}
\end{align}
with $l,\,p=$ $-L_{1}$, $1-L_{1}$, $...$, $L_{2}$, where $\operatorname*{Re}%
(x)$ denotes the real part of a complex number $x$. It is easy to show that
$\partial\mathbf{\mu/}\partial\tilde{h}_{B,p}^{\ast}=(1/2)(1+j)[x_{1-p}$,
$x_{2-p}$, $...$, $x_{N-p}]^{T}$, where $\operatorname{Im}(x)$ denotes
the imaginary part of a complex number $x$. Then, substituting this result in
(\ref{eq_9}) and keeping into account that $\partial\mathbf{R}_{\mathbf{w}%
}/\partial\tilde{h}_{l}^{\ast}=\mathbf{0}_{N}$ (where $\mathbf{0}_{N}$ denotes
the $N\times N$ null matrix) yields, after some manipulation, the expression%
\begin{align}
\left[  \mathbf{J}_{C}\left(  \mathbf{h}_{B}\right)  \right]  _{l,p} &= \frac{1}{\sigma_{w}^{2}}\operatorname*{Re}\left\{  \sum_{m=1}^{N}x_{m-l}^{\ast}x_{m-p}\right\}\nonumber\\
&= \frac{N}{\sigma_{w}^{2}}\operatorname*{Re}\left\{\frac{1}{N}\sum_{k=1-p}^{N-p}x_{k}^{\ast}x_{k+p-l}\right\}\text{.}\label{eq_12}
\end{align}
The last result shows that the FIM depends on the sample sequence $\left\{
x_{k}\right\}  $ of the channel sounding waveform $x(t)$, but is not
influenced by the parameters of the TDL channel model. We are interested in
optimizing the lower bound (\ref{eq_8bis}) (i.e., in minimizing its right hand
side) with respect to such a waveform. To tackle this optimization problem we
assume that $x(t)$ is a sample function of a \emph{bandlimited} random process
having the following properties: a) it is \emph{wide sense stationary} (WSS);
b) it has zero mean and \emph{power spectral density} (PSD) $S_{x}(f)>0$
($=0$) for $f\in(-B,B)$ ($f\notin(-B,B)$); c) its autocorrelation function
$R_{x}(\tau)$ tends to $0$ for $\tau\rightarrow\infty$ more quickly than
$1/\tau$; d) it is ergodic in autocorrelation. These assumptions entail that:
1) the sample sequence $\{x_{n}\triangleq x(t_{n})\}$ is a discrete-time WSS
random process having zero mean, autocorrelation function $R_{x}%
[l]=R_{x}(lT_{s})$ and power spectral density
\begin{align}
\overline{S}_{x}(f) &= \sum_{k=-\infty}^{+\infty} R_{x}[k] \exp\left( -j2\pi fkT_{s} \right) \nonumber\\
&=f_{s}\sum_{l=-\infty}^{+\infty}S_{x}(f-lf_{s})\text{;}\label{eq_13}
\end{align}
2) $R_{x}[l]$ decreases more quickly than $1/l$ for $l\rightarrow\infty$, so
that the series $\sum_{l=-\infty}^{\infty}\left\vert R_{x}[l]\right\vert $ is
convergent; 3) $\left\{  x_{n}\right\}  $ is ergodic in autocorrelation. Under
the above assumptions, the equality $\lim_{N\rightarrow\infty}\frac{1}{N}%
\sum_{k=1-p}^{N-p}x_{k}^{\ast}x_{k+p-l}=R_{x}[p-l]$ holds with unit
probability (see (\ref{eq_12})), so that for a finite (and large) $N$\ (i.e.,
when a large number of samples of the received signal is available for channel
estimation) the element $\left[  \mathbf{J}_{C}\left(  \mathbf{h}_{B}\right)
\right]  _{l,p}$ (\ref{eq_12}) can be approximated as%
\begin{equation}
\left[  \mathbf{J}_{C}\left(  \mathbf{h}_{B}\right)  \right]  _{l,p}\cong
\frac{N}{\sigma_{w}^{2}}R_{x}[p-l]\text{.}\label{eq_15}%
\end{equation}
The adoption of this approximation leads to a \emph{real symmetric Toeplitz}
FIM; this implies that: a) any eigenvalue of $\mathbf{J}_{C}^{-1}\left(
\mathbf{h}_{B}\right)  $ is always not smaller than $\inf\left(  \overline
{S}_{x}(f)\right)  $ \cite[lemma 4.1]{toeplitz}, so that $\operatorname{tr}%
(\mathbf{J}_{C}^{-1}\left(  \mathbf{h}_{B}\right)  )$ (see (\ref{eq_8bis}))
grows unlimitedly as $L\rightarrow\infty$ (this means that, for a given $N$,
as the number $L$ of channel parameters to be estimated increases, the overall
MSE diverges); b) the following asymptotic result holds \cite[theorem
5.2c]{toeplitz}:%
\begin{equation}
\lim_{L\rightarrow\infty}\frac{1}{L}\operatorname{tr}(\mathbf{J}_{C}%
^{-1}\left(  \mathbf{h}_{B}\right)  )=T_{s}\frac{\sigma_{w}^{2}}{N}%
\int_{-f_{s}/2}^{f_{s}/2}\frac{df}{\overline{S}_{x}(f)}\text{,}\label{eq_16}%
\end{equation}
since $N\overline{S}_{x}(f)/\sigma_{w}^{2}$ belongs to the Wiener class
(i.e., the\ sum of the absolute values of the FIM diagonal elements remains
bounded as $L\rightarrow\infty$; in other words $(N/\sigma_{w}^{2})$
$\underset{l=-\infty}{\overset{+\infty}{\sum}}\left\vert R_{xx}[l]\right\vert
<\infty$), $\overline{S}_{x}(f)\ $is a real valued function and $\overline
{S}_{x}(f)>0$ for any $f$. Then, from (\ref{eq_8bis}) and (\ref{eq_16}) the
lower bound
\begin{equation}
\lim_{L\rightarrow\infty}\frac{\varepsilon_{B,L}}{L}\geq T_{s}\frac{\sigma
_{w}^{2}}{N}\int_{-f_{s}/2}^{f_{s}/2}\frac{df}{\overline{S}_{x}(f)}%
\label{eq_17}%
\end{equation}
can be easily inferred. This result depends on the power spectrum
$\overline{S}_{x}(f)$, which can be optimized to improve the quality of
channel estimation under the constraint $T_{s}\int_{-f_{s}/2}^{f_{s}%
/2}\overline{S}_{x}(f)df=P_{x}$ on the average statistical power $P_{x}$ of
$\left\{  x_{n}\right\}  $. Applying the method of Lagrange multipliers to
this optimization problem leads to the conclusion that the right hand side of
(\ref{eq_17}) is maximised (under the given constraint) if $\overline{S}%
_{x}(f)=P_{x}$ for any $f\in(-f_{s}/2,f_{s}/2)$, i.e. if the power spectrum of
$\left\{  x_{n}\right\}  $ is \emph{uniform} (equivalently, $R_{x}%
[l]=P_{x}\delta\lbrack l]$); this occurs if (see (\ref{eq_13}))%
\begin{equation}
S_{x}(f)=\left\{
\begin{array}{ll}
\frac{\overline{S}_{x}(f)}{f_{s}}=\frac{P_{\alpha}}{f_{s}}=\frac{P_{x}}{2B} & f\in(-f_{s}/2,f_{s}/2)\\
0 & \text{elsewhere}
\end{array}
\right.  \text{,}\label{eq_18b}%
\end{equation}
since $x(t)$ is bandlimited to $f_{s}/2=B$ Hz. It is important to note that,
if the optimal power spectrum is selected for $\left\{  x_{n}\right\}  $ and
the approximation (\ref{eq_15}) is used, (\ref{eq_15}) gives $\left[
\mathbf{J}_{C}\left(  \mathbf{h}_{B}\right)  \right]  _{l,p}=(N\,P_{x}%
/\sigma_{w}^{2})\delta\lbrack p-l]$ and the FIM $\mathbf{J}_{C}\left(
\mathbf{h}_{B}\right)  $ can be put in the form%
\begin{equation}
\mathbf{J}_{C}(\mathbf{h}_{B})=\frac{N\,P_{x}}{\sigma_{w}^{2}}\mathbf{I}%
_{L}\text{,}\label{eq_19}%
\end{equation}
so that $\mathbf{J}_{C}^{-1}\left(  \mathbf{h}_{B}\right)  =(\sigma_{w}%
^{2}/(N\,P_{x}))\mathbf{I}_{L}$ and $\operatorname*{var}(\hat{h}_{l}%
)\geq\left[  \mathbf{J}_{C}^{-1}\left(  \mathbf{h}_{B}\right)  \right]
_{l,l}=\frac{1}{N}\frac{\sigma_{w}^{2}}{P_{x}}=\frac{1}{N\cdot\mathrm{SNR}}$,
where $\mathrm{SNR}\triangleq P_{x}/\sigma_{w}^{2}$ is the
\emph{signal-to-noise ratio}, and the bound (\ref{eq_8bis}) becomes
\begin{equation}
\varepsilon_{B,L}\geq\frac{L}{N\cdot\mathrm{SNR}}\triangleq\beta_{B,L}%
\text{.}\label{eq_21}%
\end{equation}
This results evidences that, for a given SNR and a given number $N$ of
processed samples, an increase in the number $L$ of significant CIR\ taps is
expected to have a negative impact on the quality of CIR estimates. Finally,
it's worth noting that the result expressed by (\ref{eq_19}) is similar to
that derived in \cite[Paragraph 3.1]{carvalho} for channel estimation based on
a training sequence that consists of \emph{a large number of uncorrelated
channel symbols}. In \cite[Paragraph 3.1]{carvalho}, however, a discrete-time
communication model is assumed in the derivation of Cramer-Rao bounds.

\subsection{BCRB-based performance limits\label{sec:BCRB}}

In this Paragraph we assume \emph{uncorrelated scattering} (US) and model the
CIR $h_{B}(t)$ as a complex Gaussian process characterized by a zero mean
(i.e., Rayleigh fading is assumed) and a PDP $P_{h}(\tau)$ with $\int
_{-\infty}^{+\infty}P_{h}(\tau)d\tau=1$. Then, we have that $\mathbf{h}_{B}$
$\sim\mathcal{C}\mathcal{N}(\mathbf{0}_{L}$, $\mathbf{R_{h}})$, where
$\mathbf{R_{h}}$ is the covariance matrix of $\mathbf{h}_{B}$; the element on
$l$-th row and $p$-th column of $\mathbf{R_{h}}$\ is given by (see (\ref{eq_4}))
\begin{gather}
\begin{array}{@{}l@{}@{}l@{}}
\mathbb{E}\left\{ h_{B,l}\, h_{B,p}^{\ast}\right\} =\mathbb{E} & \left\{ \frac{1}{2B}\int_{-B}^{B}H(f_{1})e^{j2\pi l\frac{f_{1}}{2B}}df_{1}\right.\\
 & \left.\cdot\frac{1}{2B}\int_{-B}^{B}H^{\ast}\left(f_{2}\right)e^{-j2\pi p\frac{f_{2}}{2B}}df_{2}\right\} =
\end{array}\nonumber\\
\frac{1}{\left(2B\right)^{2}}\int_{f_{2}=-B}^{B}\left[\int_{f=-B-f_{2}}^{B-f_{2}} \!\!\!\!\!\! R_{H}\left(f\right)e^{j2\pi l\frac{f}{2B}}df\right]e^{j2\pi\frac{(l-p)f_{2}}{2B}}df_{2}\label{eq_23b}
\end{gather}
with $l$, $p=-L_{1}$, $1-L_{1}$, $...$, $L_{2}$, where $R_{H}\left(  f\right)
\triangleq\mathbb{E}\left\{  H(f_{0}+f)H^{\ast}(f_{0})\right\}$ is the
\emph{channel autocorrelation function} (i.e., the inverse continuous Fourier
transform of $P_{h}(\tau)$) and $f_{0}$ is an arbitrary frequency. Note that
for $l=p$ (\ref{eq_23b}) yields
\begin{equation}\begin{split}
&\mathbb{E}\left\{  \left\vert h_{B,l}\right\vert ^{2}\right\}=\\
&\quad{}\frac{1}{\left(  2B\right)  ^{2}}\int_{f_{2}=-B}^{B}\left[  \int_{f=-B-f_{2}}^{B-f_{2}}R_{H}\left(  f\right)  e^{j2\pi l\frac{f}{2B}}df\right]  df_{2}\\
&\quad{}=\int_{y=-1/2}^{-1/2}\left[  \int_{x=-1/2-y}^{1/2-y}R_{H}\left(  2Bx\right) e^{j2\pi lx} dx\right]  dy\label{eq_23c}%
\end{split}\end{equation}
Generally speaking, channel estimation algorithms can benefit from the
availability of information about channel statistics to improve the quality of
their CIR estimate. For such algorithms a lower bound to their MSE performance
is provided by the BCRB \cite[p. 957-958]{trees}, which establishes that
$\operatorname*{MSE}(\hat{h}_{B,l})\geq\left[  \mathbf{J}_{B}^{-1}\left(
\mathbf{h}_{B}\right)  \right]  _{l,l}$ with $l=$ $-L_{1}$, $1-L_{1}$, $...$,
$L_{2}$, where $\mathbf{J}_{B}\left(  \mathbf{h}_{B}\right)  $ is an $L\times
L$ complex matrix, known as \emph{Bayesian} \emph{Fisher Information Matrix}
(BFIM). The element on the $l$-th row and $p$-th column of $\mathbf{J}%
_{B}\left(  \mathbf{h}\right)  $ can be evaluated as \cite[equ. 53]{dong}
\begin{equation}
\left[  \mathbf{J}_{B}(\mathbf{h}_{B}\mathbf{)}\right]  _{l,p}=\left[
\mathbf{J}_{C}(\mathbf{h}_{B})\right]  _{l,p}+\left[  \mathbf{J}%
_{h}(\mathbf{h}_{B})\right]  _{l,p}\label{eq:bcrb_def}%
\end{equation}
where $\mathbf{J}_{C}(\mathbf{h}_{B})$ is the CRB FIM evaluated in the
previous Paragraph and
\begin{equation}\begin{split}
& \left[\mathbf{J}_{h}(\mathbf{h}_{B})\right] _{l,p}\triangleq \\
&\quad{}\mathbb{E}_{\mathbf{h}_{B}}  \left. \left\{  \frac{\partial\ln f_{\mathbf{h}_{B}}\left(  \mathbf{\tilde{h}}_{B}\right)  }{\partial\tilde{h}_{B,l}^{\ast}}  \left(  \frac{\partial\ln f_{\mathbf{h}_{B}}\left(  \mathbf{\tilde{h}}_{B}\right)  }{\partial\tilde{h}_{B,p}^{\ast}}\right)  ^{\ast}\right\} \right\vert _{\mathbf{\tilde{h}}_{B}=\mathbf{h}_{B}}  \text{.}\label{eq_25}
\end{split}\end{equation}
where $f_{\mathbf{h}_{B}}\left(  \mathbf{\tilde{h}}_{B}\right)  $ denotes the
joint pdf of $\mathbf{h}_{B}$. Like in the previous case (see (\ref{eq:mse})
and (\ref{eq_8bis})) the bound
\begin{equation}\begin{split}
\bar{\varepsilon}_{B,L} &= \sum_{l=-L_{1}}^{L_{2}}\mathbb{E}_{w,h_{B,l}%
}\left\{  \left\vert h_{B,l}-\hat{h}_{B,l}\right\vert ^{2}\right\}  \geq \\
& \sum_{l=-L_{1}}^{L_{2}}\left[  \mathbf{J}_{B}^{-1}\left(  \mathbf{h}%
_{B}\right)  \right]  _{l,l}=\operatorname*{tr}\left(  \mathbf{J}_{B}%
^{-1}(\mathbf{h}_{B})\right)  \triangleq\bar{\beta}_{B,L}\label{eq_25b}%
\end{split}\end{equation}
can easily be developed for $\bar{\varepsilon}_{B,L}$ (\ref{MSEbis}). To
evaluate the right hand side of the last inequality, let us compute now the
partial derivatives appearing in (\ref{eq_25}). It is easy to show that
\begin{equation}\begin{split}
&\frac{\partial\ln f_{\mathbf{h}_{B}}\left(  \mathbf{\tilde{h}}_{B}\right)}{\partial\tilde{h}_{B,p}^{\ast}}=\\
&\quad{}-\frac{1}{2}\left(  \frac{\partial\left(\mathbf{\tilde{h}}_{B}^{H}\mathbf{R_{h}^{-1}\tilde{h}}_{B}\right)}{\partial\mathrm{Re}\left\{  \tilde{h}_{B,p}\right\}  }+j\frac{\partial\left(\mathbf{\tilde{h}}_{B}^{H}\mathbf{R_{h}^{-1}\tilde{h}}_{B}\right)  }{\partial\mathrm{Im}\left\{  \tilde{h}_{B,p}\right\}  }\right)  =\\
&\quad{}-\left[\mathbf{R_{h}^{-1}}\mathbf{\tilde{h}}_{B}\right]  _{p}\text{.}\label{eq_29}
\end{split}\end{equation}
Then, substituting (\ref{eq_29}) in (\ref{eq_25}) yields
\begin{align}
\mathbf{J}_{h}(\mathbf{h}_{B})&=\mathbb{E}_{\mathbf{h}_{B}}\left\{  \left.\left(  \mathbf{R_{h}^{-1}}\mathbf{\tilde{h}}_{B}\right)  \left(\mathbf{R_{h}^{-1}}\mathbf{\tilde{h}}_{B}\right)  ^{H}\right\vert_{\mathbf{\tilde{h}}_{B}=\mathbf{h}_{B}}\right\} \nonumber\\
&=\left(  \mathbf{R}_{\mathbf{h}}^{-1}\right)  ^{H}=\mathbf{R}_{\mathbf{h}}^{-1}\text{,}\label{eq_30}
\end{align}
since $\mathbf{R_{h}}$ is an Hermitian matrix. Like the CRB, the BCRB is
influenced by the choice of the sounding waveform through $\mathbf{J}_{C}(\mathbf{h}_{B})$ (see (\ref{eq:bcrb_def})); in the following a uniform power spectrum is assumed for this waveform (see (\ref{eq_18b})). Then, substituting (\ref{eq_19}) and (\ref{eq_30}) in (\ref{eq:bcrb_def}) yields
\begin{align}
\mathbf{J}_{B}(\mathbf{h}_{B}) &= \frac{N\,P_{x}}{\sigma_{w}^{2}}\mathbf{I}_{L}+\mathbf{R_{h}^{-1}}\nonumber\\
&=N\,\cdot\mathrm{SNR}\left(  \mathbf{I}_{L}+\frac{\mathbf{R_{h}^{-1}}}{N\cdot\mathrm{SNR}}\right)  \text{.}\label{eq:bcrb_opt}
\end{align}
Unluckily, $\mathbf{J}_{B}(\mathbf{h}_{B})$ is not a Toepliz matrix and, as
far as we know, no asymptotic result is available for the trace of its
inverse. However, a simple expression for this trace can be derived if the
Taylor series representation%
\begin{equation}
\mathbf{J}_{B}^{-1}(\mathbf{h}_{B})=\frac{1}{N\cdot\mathrm{SNR}}\sum_{k=0}^{\infty}\left(  -\frac{\mathbf{R_{h}^{-1}}}{N\cdot\mathrm{SNR}}\right)  ^{k}\label{eq_31}
\end{equation}
can be adopted for $\mathbf{J}_{B}^{-1}(\mathbf{h}_{B})$; this holds if the
$L$ eigenvalues of the matrix $(1/(N\cdot\mathrm{SNR}))\mathbf{R_{h}^{-1}}$
are distinct and their values are less than unity\footnote{The eigenvalues of
the covariance matrix $\mathbf{R_{h}}$ are always positive;\ this implies that
the eigenvalues of the matrix $\mathbf{R_{h}^{-1}}$ are also positive.}, i.e.
$1/(N\cdot\mathrm{SNR}\cdot\lambda_{i})<1$ (or, equivalently, $\ \lambda
_{i}>\frac{1}{N\cdot\mathrm{SNR}}>0$) for $i=1$, $2$, $...$, $L$, where
$\left\{  \lambda_{i}\text{, }i=1\text{, }2\text{, }...\text{, }L\right\}  $
denote the (real) eigenvalues of $\mathbf{R_{h}}$. In fact, this
representation entails that
\begin{equation}
\operatorname*{tr}\left\{  \mathbf{J}_{B}^{-1}(\mathbf{h}_{B})\right\}=\frac{1}{N\cdot\mathrm{SNR}}\sum_{k=0}^{\infty}\operatorname*{tr}\left\{  \left(  -\frac{\mathbf{R_{h}^{-1}}}{N\cdot\mathrm{SNR}}\right)^{k}\right\}  \text{.}\label{eq_32}%
\end{equation}
Since $\mathbf{R_{h}}$ is an hermitian matrix, its inverse $\mathbf{R_{h}%
^{-1}}$ can be factored as $\mathbf{R_{h}^{-1}}=\mathbf{U\,}\mathbf{\Sigma
^{-1}}\mathbf{U}^{H}$ \cite[p. 245, sec. 5.2]{strang}, where $\mathbf{U}$ is a
$L\times L$ unitary matrix (whose columns are the eigenvectors of
$\mathbf{R_{h}}$) and $\mathbf{\Sigma}=\operatorname*{diag}\left\{
\lambda_{1}\text{, }\lambda_{2}\text{, }...\text{, }\lambda_{L}\right\}  $.
Exploiting this factorisation it can be easily shown that
\begin{align}
\operatorname*{tr}\left\{  \left(  -\frac{\mathbf{R_{h}^{-1}}}{N\cdot\mathrm{SNR}}\right)  ^{k}\right\}&= \operatorname*{tr}\left\{  \left( \frac{-1}{N\cdot\mathrm{SNR}}\mathbf{\Sigma^{-1}}\right)  ^{k}\right\}\nonumber\\
&= \left(  \frac{-1}{N\cdot\mathrm{SNR}}\right)  ^{k}\sum_{i=1}^{L}\frac{1}{\lambda_{i}^{k}}\label{eq_35}
\end{align}
since $\operatorname*{tr}\left\{  \mathbf{U\,D\,U}^{H}\right\}
=\operatorname*{tr}\left\{  \mathbf{D}\right\}  $ for any matrix
$\mathbf{D}$ (this result is known as \emph{similarity invariance property} of
the trace operator). Then, substituting the last result in (\ref{eq_32})
yields
\begin{align}
\operatorname*{tr}\left(  \mathbf{J}_{B}^{-1}(\mathbf{h}_{B})\right)   &=\frac{1}{N\cdot\mathrm{SNR}}\sum_{k=0}^{\infty}\sum_{i=1}^{L}\left(  \frac{-1}{N\cdot\mathrm{SNR}\cdot\lambda_{i}}\right)^{k}\nonumber\\
&=\frac{1}{N\cdot\mathrm{SNR}}\sum_{i=1}^{L}\frac{1}{1+\frac{1}{N\cdot\mathrm{SNR}\cdot\lambda_{i}}}\nonumber\\
&=\sum_{i=1}^{L}\frac{1}{N\cdot\mathrm{SNR}+\frac{1}{\lambda_{i}}}\text{,}\label{eq_36}
\end{align}
since we have assumed that $1/(N\cdot\mathrm{SNR}\cdot\lambda_{i})<1$ for
$i=1$, $2$, $...$, $L$. Finally, substituting (\ref{eq_36}) in (\ref{eq_25b})
yields the bound%
\begin{equation}
\bar{\varepsilon}_{B,L}\geq\sum_{i=1}^{L}\frac{1}{N\cdot\mathrm{SNR}+\frac
{1}{\lambda_{i}}}\triangleq\bar{\beta}_{B,L}\text{.}\label{eq_37}%
\end{equation}
It is worth noting that this bound depends on the statistical properties of
the channel through the eigenvalues of the matrix $\mathbf{R_{h}}$, whose
structure is related to the shape of $R_{H}\left(  f\right)  $ (or,
equivalently, of $P_{h}(\tau)$). Let us try now to simplify this bound under
the assumption that the bandwidth $B$ of the sounding signal is substantially
larger than the coherence bandwidth $B_{c}$\ of the communication channel
(\emph{wideband channel sounding}). In this case we have that\footnote{This
approximation is motivated by the fact that $B_{c}$ provides an indication of
the width of $R_{H}(f)$ (i.e., of the frequency interval over which $R_{H}(f)$
takes on significant values). Then, if $B\gg B_{c}$, the following integral is
negligibly influenced by a change in the center ($f_{2} $) of the integration
interval.} (see (\ref{eq_23b})) $\int_{f=-B-f_{2}}^{B-f_{2}}R_{H}\left(
f\right)  \exp\left(  j2\pi l\frac{f}{2B}\right)  df\cong P_{h}\left(
\frac{l}{2B}\right)  \cong P_{h}(0)$ for any $f_{2}\in(-B,B)$, so that
$\mathbb{E}\left\{  h_{B,l}\,h_{B,k}^{\ast}\right\}  \cong P_{h}(0)/2B$ if
$l=k$ and $=0$ if $l\neq k$. Then, the channel taps are uncorrelated,
$\mathbf{R_{h}^{-1}}=(2B/P_{h}(0))\mathbf{I}_{L}$, and (see (\ref{eq:bcrb_opt}%
)) $\mathbf{J}_{B}(\mathbf{h}_{B})=\left(  N\cdot\mathrm{SNR}+\frac{2B}%
{P_{h}(0)}\right)  \mathbf{I}_{L}$, so that the bound (\ref{eq_25b}) becomes%
\begin{equation}
\bar{\varepsilon}_{B,L}\geq\frac{L}{N\cdot\mathrm{SNR}+2B/P_{h}%
(0)}\triangleq\bar{\beta}_{B,L}^{(w)}\text{.}\label{eq_40}%
\end{equation}
Note that $2B/P_{h}(0)\gg1$ because of the assumption of wideband signalling
over the communication channel. Therefore, a comparison of the last result
with (\ref{eq_21}) evidences that, in this scenario, a significant improvement
in the quality of channel estimates should be expected if the channel
estimator is endowed with a knowledge of the channel statistics.

Finally, we note that the result (\ref{eq_40}) is substantially different from
the BCRB\ evaluated in \cite[Appendix A]{dong}, which refers to a
discrete-time channel model in which the channel taps are independent and
identically distributed random variables with a given pdf.

\section{Numerical Results\label{sec:Numerical-results}}

\begin{table}[tp] \centering
\begin{tabular}
[c]{|l|l|l|l|l|}\hline
& E & G & U & TE\\\hline
$B=1/\tau_{ds}$ & $(1,5)$ & (3,4) & (1,6) & (1,6)\\\hline
$B=10/\tau_{ds}$ & (1,48) & (33,33) & (1,61) & (1,63)\\\hline
\end{tabular}
\caption{Values of the couple $(L_{1}, L_{2})$  capturing at least 90\% of
the overall average energy of $h_{B}(t)$\label{tabella}.}
\end{table}

\begin{figure}[tp]
\centering \includegraphics[width=3in]{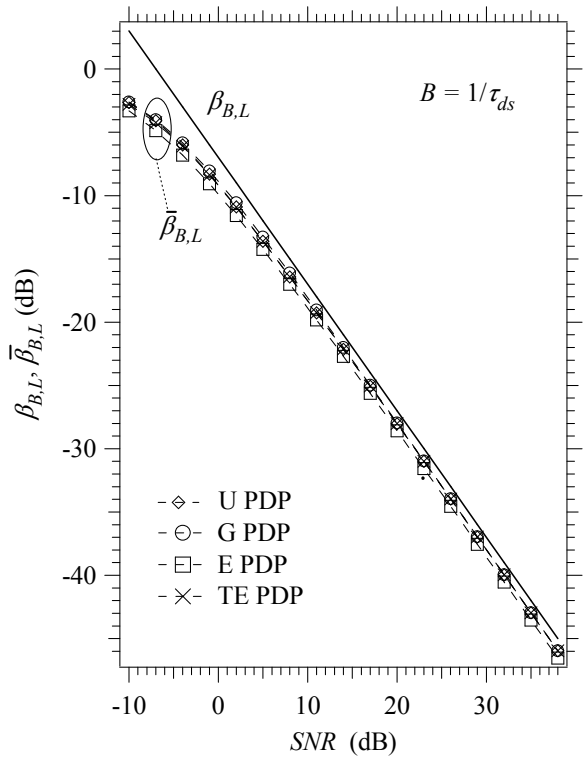} 
\caption{Performance bounds $\beta_{B,L}$ (\ref{eq_21}) and $\bar{\beta}_{B,L}$ (\ref{eq_25b}) versus the SNR for different PDP's in the case $B=1/\tau_{ds}$.}\label{fig:crb_bcrb_narrowband}
\end{figure}

\begin{figure}[tp]
\centering \includegraphics[width=3in]{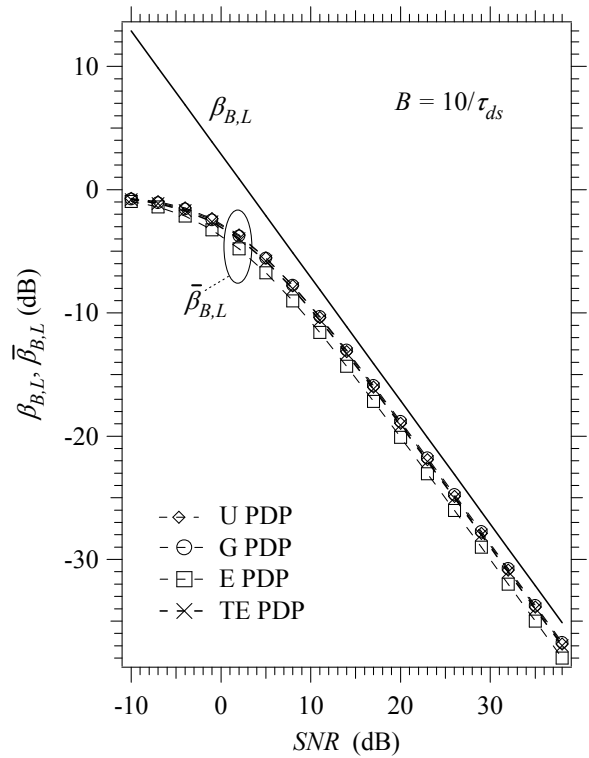} 
\caption{Performance bounds $\beta_{B,L}$ (\ref{eq_21}) and $\bar{\beta}_{B,L}$ (\ref{eq_25b}) versus the SNR for different PDP's in the case $B=10/\tau_{ds}$.}\label{fig:crb_bcrb_wideband}
\end{figure}

The bounds expressed by (\ref{eq_21}) and (\ref{eq_25b}) (with $\mathbf{J}%
_{B}(\mathbf{h})$ given by (\ref{eq:bcrb_opt})) have been evaluated for an
\emph{exponential} (E), a \emph{Gaussian} (G), a \emph{uniform} (U) and a
\emph{truncated exponential} (TE) PDP \cite{chiavaccini}, so that 
$P_{h}(\tau)=\frac{e^{-\tau/\tau_{ds}}}{\tau_{ds}}\operatorname*{u}(\tau)$,
$P_{h}(\tau)=\frac{e^{-\tau^{2}/(2\tau_{ds}^{2})}}{\tau_{ds}\sqrt{2\pi}}$,
$P_{h}(\tau)=\frac{\operatorname*{u}(\tau)-\operatorname*{u}(\tau-\tau_{ds}\sqrt{12})}{\tau_{ds}\sqrt{12}}$,
$P_{h}(\tau)=\frac{\operatorname*{u}(\tau)-\operatorname*{u}(\tau-\tau_{M})}{\tau_{0}(1-e^{-\tau_{M}/\tau_{0}})}e^{-\tau/\tau_{0}}$ respectively, where $\operatorname*{u}(\tau)$ is the unitary step function, $\tau_{ds}$ is the \emph{rms channel
delay spread}, $\tau_{M}$ is the maximum delay in the TE PDP and $\tau_{0}$ is
another time parameter depending on $\tau_{ds}$ (see \cite[eq. (16)]%
{chiavaccini}). In our simulations the channel bandwidths $B=1/\tau_{ds}$ and
$B=10/\tau_{ds}$ (wideband channel sounding) have been taken into
consideration. In both cases and for each of the above mentioned PDP's we have
evaluated the smallest values of the parameters $L_{1}$ and $L_{2}$ ensuring
that the overall average energy $\sum_{l=-L_{1}}^{L_{2}}\mathbb{E}\{\left\vert
h_{B,l}\right\vert ^{2}\}$ (where $\mathbb{E}\{\left\vert h_{B,l}\right\vert
^{2}\}$ is given by (\ref{eq_23c})) associated with the RHS of (\ref{eq_3}) is
at least $90\%$ of the overall average energy of $h_{B}\left(  t\right)  $
(see Table \ref{tabella}). Then, on the basis of such values, the couples
$(L_{1},L_{2})=(3,6)$ and $(L_{1},L_{2})=(33,63)$\ have been selected for
$B=1/\tau_{ds}$ and $B=10/\tau_{ds}$, respectively, since they encompass all
the cases of Table \ref{tabella}. Fig. \ref{fig:crb_bcrb_narrowband} (Fig.
\ref{fig:crb_bcrb_wideband}) illustrates the bounds $\beta_{B,L}$
(\ref{eq_21}) and $\bar{\beta}_{B,L}$ (\ref{eq_25b}) versus the SNR for
$B=1/\tau_{ds}$ ($B=10/\tau_{ds}$)\ and all the considered PDP's. These
results show that: a) independently of the bandwidth adopted for data
transmission, the impact of the availability of a priori information on the
estimation accuracy of a communication channel is significant mainly at low
SNR's (where the terms $\{1/\lambda_{i}\}$ (\ref{eq_37}), not included in
(\ref{eq_21}), yield a performance floor);\ b) the BCRB\ is negligibly
influenced by the PDP type; c) there is a significant performance gap between
the case $B=10/\tau_{ds}$ and $B=1/\tau_{ds}$ (this is due to the fact that
the overall number of channel taps to be estimated in the latter case is
substantially smaller than that of the former one). Our simulations have also
evidenced that: 1) in the considered scenarios an accurate approximation of
(\ref{eq_25b}) is provided by eq. (\ref{eq_37})\ for both values of $B$; 2)
eq. (\ref{eq_40}) represents a loose bound for the case $B=10/\tau_{ds}$.

\section{Conclusions\label{sec:conclusions}}

The problem of assessing performance limits on pilot-aided channel estimation
of a time-continuous frequency selective channel has been investigated. Novel
bounds based on the CRB and the BCRB for TDL channel models have been derived
and have been assessed for two different scenarios. The derived results shed
new light on the achievable limits of pilot-aided channel estimation and the
properties of optimal waveforms for channel sounding.

\end{document}